\documentclass[aps,prd,twocolumn,groupedaddress,showpacs,nofootinbib]{revtex4}

\usepackage{graphicx}

\begin{document}


\title{Appearance of neutronization peak and decaying supernova
neutrinos}

\author{Shin'ichiro Ando}
\email[Email address: ]{ando@utap.phys.s.u-tokyo.ac.jp}
\affiliation{Department of Physics, School of Science, The University of
Tokyo, 7-3-1 Hongo, Bunkyo-ku, Tokyo 113-0033, Japan}

\date{Submitted April 13, 2004; accepted May 20, 2004}

\begin{abstract}

Nonradiative neutrino decay, which is not satisfactorily constrained,
 possibly and significantly changes the detected neutrino signal from
 galactic supernova explosions.
We focus on the appearance of a sharp peak due to a neutronization burst
 in the time profile; this phenomenon would occur if the original
 $\nu_e$, produced at the neutrinosphere and becoming $\nu_2$ or $\nu_3$
 at the stellar surface, decays into a lighter antineutrino state such
 as $\bar\nu_1$ or $\bar\nu_2$ coupled to $\bar\nu_e$.
This is due to the fact that the signature of the neutronization burst
 is common to all numerical simulations, contrary to the spectral energy
 distribution of each flavor neutrino and antineutrino, which is still
 under intense debate.
Therefore, the appearance of a neutronization peak in the $\bar\nu_e$
 signal, if it were detected, would clearly indicate the nonstandard
 properties of neutrinos; the nonradiative neutrino decay would be one
 of the possible candidates.
Using a newly developed formulation that includes flavor conversions
 inside the supernova envelope and neutrino decay during propagation in
 a vacuum, we calculate the expected neutrino signal at the detectors;
 the lifetimes of three modes $\tau_{12}$, $\tau_{13}$, and $\tau_{23}$
 are taken to be free parameters.
We further introduce simple quantities, which represent a peak sharpness
 of the time profile and spectral hardness, and investigate the
 parameter dependence of these quantities.
As the result, it is found that they are quite dependent on the relevant
 parameters, but it would be quite difficult to distinguish models using
 the signal obtained by the Super-Kamiokande detector; the future
 megaton-class detectors would have sufficient sensitivity.
We also compare the neutrino decay model with another mechanism---i.e.,
 resonant spin-flavor conversion---which also may give the appearance of
 a neutronization peak, and conclude that these two independent
 mechanisms give a very different signal and one can be distinguishable
 from the other.

\end{abstract}


\maketitle

\section{Introduction \label{sec:Introduction}}

In recent years, we have made great progress concerning our knowledge of
neutrino properties; especially, many ground-based experiments, which
observed atmospheric \citep{Fukuda99}, solar
\citep{Fukuda02a}, and reactor neutrinos \citep{Eguchi03a}, have
revealed nonzero neutrino masses and mixing angles---i.e., properties
beyond the standard model of particle physics.
However, many other neutrino properties are left unknown, such as the
nonzero magnetic moment and neutrino decay.
Fortunately, our current knowledge of the mass differences as well as
mixing angles enables us to consider these further exotic properties.

The most stringent and precise limits on both the neutrino magnetic
moment and lifetime of nonradiative neutrino decay are obtained by solar
neutrino observations.
The basic technique for both cases is as follows.
We already know that the famous solar neutrino problem is best explained
by the large mixing angle (LMA) solution.
If other mechanisms, such as the magnetic moment or nonradiative decay,
work in nature, it should change the observed signal slightly; i.e., no
positive claim can then be used to put a limit on the other exotic
properties.
For example, the neutrino magnetic moment, if ever, would produce
$\bar\nu_e$ via spin-flavor precession inside the Sun
\citep{Lim88,Lim90}.
However, the recent KamLAND experiment report that no $\bar\nu_e$
candidates were found from the Sun and this result is used to obtain the
upper limit $\mu_\nu B_T(0.05R_\odot)\alt 10^{-5}\mu_B~{\rm G}$, where
$\mu_B$ is the Bohr magneton and $B_T$ represents the transverse
component of the solar magnetic field \citep{Eguchi03b}.
As for nonradiative neutrino decay, the lower limit to the lifetime is
obtained to be $\tau_{12}/m\agt 10^{-4}$ s/eV, owing to no positive
signature of the decay in the flux and spectrum of the solar neutrinos
\citep{Beacom02} (see also Ref. \citep{Eguchi03b}, for a more recent and
stringent limit using the KamLAND data).
For other laboraory bounds, we refer the reader to
Refs. \citep{Daraktchieva03,Barger82}.

These current limits are, however, still rather weak and such exotic
mechanisms we consider potentially work in a more extreme environment as
actually expected in the case of core collapse supernovae.
Spin-flavor conversion of supernova neutrinos, induced by the
interaction between the neutrino magnetic moment and the supernova
magnetic fields, has been studied by many researchers
\citep{Lim88,Voloshin88,Ando03g,Akhmedov03b}
and found to give a leading effect on the neutrino spectrum and
luminosity curve detected at the Earth.
Also, in the case of neutrino decay, because galactic supernovae are
expected to be located at a much more distant place compared with the
Sun, a far more stringent lower limit to the lifetime is expected.
In addition, it has recently been proposed that the diffuse background
of neutrinos emitted by past supernova explosions may be used to probe
the lifetime of neutrinos \citep{Ando03f}.
Thus, core-collapse supernovae are considered to be a wonderful
astrophysical event, which can also be used as a laboratory for particle
physics beyond the standard model.
Other high-energy astrophysical objects are also expected to be
available for this purpose \citep{Beacom03}.

However, the expected galactic supernova neutrino signal including
nonradiative decay has not been discussed precisely; only very rough
discussions have been done.
Therefore, in the present paper, we give comprehensively the expected
supernova neutrino signal at the large volume water \v Cerenkov
detectors on the Earth using realistic models of original neutrino
spectrum and luminosity curve numerically calculated by
\citet{Thompson03}.
In particular, we focus on the case in which the sharp peak of a
neutronization burst appears in the detected $\bar\nu_e$ signal at the
water \v Cerenkov detectors.
This is due to the original $\nu_e$, produced by neutronization of the
protoneutron star matter, possibly decays into $\bar\nu_e$ component
during its propagation.
In fact, in the previous paper we have pointed out that spin-flavor
conversions in the supernova envelope can also cause the same
phenomenon---i.e., the appearance of a neutronization peak in the
$\bar\nu_e$ signal \citep{Ando03g} (see also Ref. \citep{Akhmedov03b}).
Hence, we compare the expected neutronization peak due to neutrino decay
with that due to spin-flavor conversion and discuss their difference.
Anyway, the appearance of such a signature clearly indicates a
nonstandard neutrino property; in that case, neutrino decay would be one
of the possible candidates.
On the other hand, the obtained neutrino spectrum would be useful, but
the shape of the original spectrum is still matter of debate.
Although the difference among several groups is not very large, such a
slight difference gives a large uncertainty when we discuss flavor
conversion mechanisms or decay during propagation; i.e., whether the
observed signature comes from the intrinsic or extrinsic effect (such as
decay) would be quite unknown at present.

It should be noted that in this study we consider only vacuum neutrino
decays.
It is possible, however, to construct models where fast invisible decays
can be triggered by matter effects
\citep{Berezhiani87,Berezhiani92,Kachelriess00,Farzan03}
at the very high densities characterizing the supernova neutrinosphere,
even in the absence of vacuum decays.
In such scenarios, matter-induced decays (or interactions) might thus
occur before flavor conversions in supernovae, leading to a
phenomenology rather different from the one considered in this paper.
We emphasize that the results discussed in the following sections are
generally applicable to vacuum neutrino decay occurring after flavor
transitions and our approach is not constrained from the supernova
cooling discussion as detailed later.
In addition, with the lifetime considered in this study, the flavor
conversions occur well before the decay and mass eigenstates are
expected to become incoherent.
Thus, the interference effects between the two phenomena as discussed in
Ref. \citep{Lindner01} can be neglected in our treatment.

This paper is organized as follows.
In Sec.~\ref{sec:Models of decaying neutrinos}, we give several
descriptions of models of decaying neutrinos and introduce a specific
formulation of the decay rate from the Lagrangian.
In Sec.~\ref{sec:Original supernova neutrino signal}, an adopted model
for the original (which means {\it before} occurring extrinsic processes
such as flavor conversions or decays during propagation) spectrum and
luminosity curve of supernova neutrinos are introduced; the effect of
flavor conversions (without decay) is also described briefly.
A newly developed formulation including both flavor conversions and
decay is presented in Sec.~\ref{sec:Formulation}, and we further give
the calculated results obtained with the formulation in
Sec.~\ref{sec:Results}.
Finally, we discuss several other possibilities in
Sec.~\ref{sec:Discussion} and a brief summary is given in
Sec.~\ref{sec:Conclusions}.

\section{Models of decaying neutrinos \label{sec:Models of decaying
neutrinos}}

In this paper, we study nonradiative two-body decay of the ``invisible
mode''---i.e., decays into possibly detectable neutrinos plus truly
invisible particles---e.g., light scalar or pseudoscalar bosons such as
the Majoron \citep{Chikashige81}.
On the other hand, radiative neutrino decay $\nu_j\to\nu_i\gamma$ has
been considered in many papers (see Ref. \citep{Raffelt96} and
references therein) and very stringent limits on the lifetime-to-mass
ratio, $\tau /m\agt 10^{20}$ s/eV, have already been set by several
arguments \citep{Ressel90}.
The Majoron models that cause nonradiative neutrino decay typically have
tree-level scalar or pseudoscalar couplings of the form
\begin{equation}
 \mathcal L=g_{ij}\bar\nu_i\nu_j\chi
  +h_{ij}\bar\nu_j\gamma_5\nu_j\chi +{\rm H.c.},
  \label{eq:Lagrangian}
\end{equation}
where $\chi$ represents a massless Majoron, which does not carry a
definite lepton number.
For the coupling specified by Eq. (\ref{eq:Lagrangian}), the decay rates
into neutrino and antineutrino daughters are given by
\citep{Kim90,Berezhiani92}
\begin{eqnarray}
 {\mit\Gamma}_{\nu_2\to\nu_1}&=&\frac{m_1m_2}{16\pi E_2}
  \left[g^2\left(\frac{x}{2}+2+\frac{2}{x}\ln x-\frac{2}{x^2}
	  -\frac{1}{2x^3}\right)\right.\nonumber\\
 &&{}\left.+h^2\left(\frac{x}{2}-2+\frac{2}{x}\ln x+\frac{2}{x^2}
	      -\frac{1}{2x^3}\right)\right],
 \label{eq:Gamma_12 for conserved case}\\
  {\mit\Gamma}_{\nu_2\to\bar\nu_1}&=&\frac{m_1m_2}{16\pi E_2}
  \left(g^2+h^2\right)
  \left[\frac{x}{2}-\frac{2}{x}\ln x-\frac{1}{2x^3}\right],
  \label{eq:Gamma_12 for flipped case}
\end{eqnarray}
where $x=m_2/m_1$, and we have dropped the subscripts on the coupling
constants.
Analogous expressions hold for $\bar\nu$ decay, with the replacements
$\nu\leftrightarrow\bar\nu$.
The decay widths in this paper are defined in the laboratory frame, and
therefore the relation to the rest-frame lifetimes is ${\mit\Gamma}
(E)=m/(\tau E)$.
From these two expressions (\ref{eq:Gamma_12 for conserved case}) and
(\ref{eq:Gamma_12 for flipped case}) it is easily seen that the decay
rate is dependent on whether the helicity flips or conserves as well as
whether the neutrino masses are quasidegenerate ($x\approx 1$) or
hierarchical ($x\gg 1$).
In the case of strongly hierarchical masses, we obtain
${\mit\Gamma}_{\nu_2\to\nu_1}\approx {\mit\Gamma}_{\nu_2\to\bar\nu_1}$;
on the other hand, in the case of quasidegenerate masses,
Eqs. (\ref{eq:Gamma_12 for conserved case}) and (\ref{eq:Gamma_12 for
flipped case}) lead to the relation ${\mit\Gamma}_{\nu_2\to\nu_1}\gg
{\mit\Gamma}_{\nu_2\to\bar\nu_1}$, unless the coupling constant $g$ is
strongly suppressed compared with $h$.
Therefore, if the neutrino masses are quasidegenerate, the produced
neutrinos decay into neutrinos dominantly (helicity-conserved channel),
while little into antineutrinos (helicity-flipped channel); hierarchical
masses allow both decay channels by the same branching ratio.

The distribution of the energies of daughter (anti)neutrinos is very
dependent on whether the masses are hierarchical or quasidegenerate.
In the former case, the energy of daughter neutrinos is generally
degraded, and its distribution is given by
\begin{equation}
\psi_c(E_p,E_d) = \frac{2E_d}{E_p^2},
~~\psi_f(E_p,E_d) = \frac{2}{E_p}\left(1-\frac{E_d}{E_p}\right),
\label{eq:energy distribution of daughter neutrinos}
\end{equation}
where $E_p$ and $E_d$ represent the energy of parent and daughter
neutrinos, respectively, and $\psi_c$ and $\psi_f$ are the distribution
function of the helicity-conserved and helicity-flipped channels,
respectively.
In the latter case, on the other hand, the daughter neutrino energy is
almost the same as that of parent neutrinos; the energy distribution
becomes a delta function $\delta (E_p-E_d)$.

As already mentioned in Sec.~\ref{sec:Introduction}, we are mainly
interested in the appearance of a neutronization peak at detectors,
which dominantly catches $\bar\nu_e$; this is because the signature of
the neutronization burst is very common to any supernova simulations,
and its detection in the $\bar\nu_e$ channel would be a smoking gun to
the nonstandard properties of neutrinos.
Thus, the case of quasidegenerate masses, which prohibits the
helicity-flipped channel---i.e., $\nu_e\to\bar\nu_e$, is not our prime
interest.
Although the obtained spectrum would also be helpful even in that case,
we assume that the neutrino masses are strongly hierarchical ($m_1\ll
m_2\ll m_3$) from this point on.

At the end of this section, we mention the constraints on the coupling
constants $g$ and $h$ from the supernova cooling argument.
In the dense core of supernovae, contrary to the decay in vacuum, the
Majoron cooling proceeds mainly via $\nu_e\nu_e\to\chi$.
A conservative upper limit on the coupling constant $|g_{ee}|$ is
obtained by the fact that the luminosity of the Majoron, $L_\chi$,
should not exceed $3\times 10^{53}$ erg s${}^{-1}$ \citep{Farzan03},
because we already know that almost all the gravitational binding energy
of new-born neutron stars is released as neutrinos by the actual
detection of supernova neutrinos by Kamiokande \citep{Hirata87} and IMB
\citep{Bionta87} detectors.
This discussion translates into the bound $|g_{ee}|\alt 4\times
10^{-7}$ and it is the strongest constraint on the parameter compared
with other experimental ones such as of neutrinoless double-beta decay
with Majoron emission \citep{Zuber98}.
Bounds on other parameters such as $|g_{e\mu}|$ and $|g_{\mu\mu}|$ are
also obtained by the similar arguments (for a detailed discussion, see,
e.g., Ref. \citep{Farzan03}).
Our discussion in the present paper, however, is completely free of such
stringent constraints.
This is because the relevant parameters in our case are the coupling
constants in the basis of mass eigenstates $g_{ij}$, while those in the
cooling argument are in the basis of flavor eigenstates
$g_{\alpha\beta}$.
These two expressions in different bases are related to each other as
\begin{equation}
g_{ij}=\sum_{\alpha\beta}g_{\alpha\beta}U_{\alpha i}^\ast U_{\beta j}.
\label{eq:relation between g's}
\end{equation}
Since the mixing matrix $U_{\alpha i}$ contains several unknown
parameters such as $\theta_{13}$ or the $CP$-violating phase $\delta$,
the limits from supernova cooling do not directly translate into
$g_{ij}$ relevant for our study.

\section{Original supernova neutrino signal and flavor conversions
\label{sec:Original supernova neutrino signal}}

\begin{figure}[htbp]
\begin{center}
\includegraphics[width=8cm]{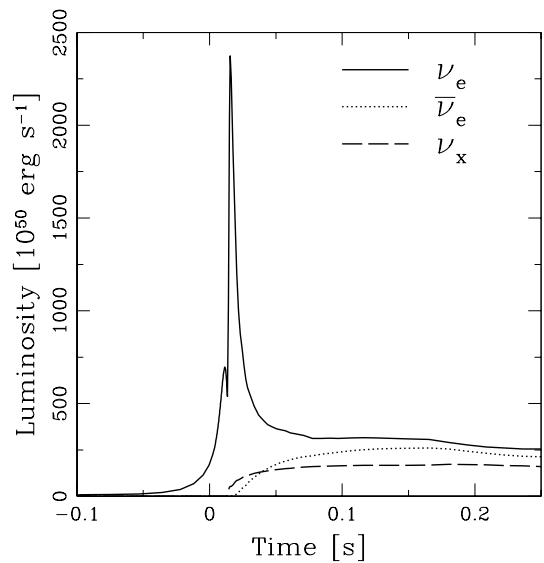}
\caption{Original luminosity of the emitted neutrinos as a function of
 time, calculated by \citet{Thompson03}. The progenitor mass is
 $15M_\odot$. \label{fig:luminosity_curve_org}}
\end{center}
\end{figure}

\begin{figure}[htbp]
\begin{center}
\includegraphics[width=8cm]{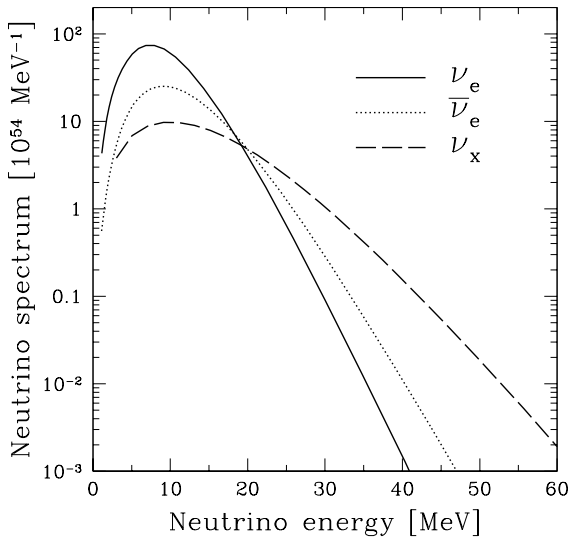}
\caption{Original neutrino spectrum integrated to 0.25 s after core
 bounce, calculated by \citet{Thompson03}. The progenitor mass is
 $15M_\odot$. \label{fig:spectrum_org}}
\end{center}
\end{figure}

We adopt as the original neutrino spectrum as well as the luminosity
curve, the results of the numerical simulation by \citet{Thompson03}; we
use the model calculated for the 15$M_\odot$ progenitor star.
Their calculation has particularly focused on shock breakout and
followed the dynamical evolution of the cores through collapse until the
first 250 ms after core bounce.
They have incorporated all the relevant neutrino processes such as
neutrino-nucleon scatterings with nucleon recoil as well as nucleon
bremsstrahlung; these reactions have recently been recognized to give a
non-negligible contribution to the spectral formation.
In Figs. \ref{fig:luminosity_curve_org} and \ref{fig:spectrum_org}, we
show the original luminosity curve and number spectrum of neutrinos,
respectively.
In these figures, $\nu_x$ represents nonelectron neutrinos and
antineutrinos.

The neutrino luminosity curve is quite characteristic among the
different flavors.
In particular, there is a very sharp peak of $\nu_e$ called the
neutronization burst, whose duration is typically $\sim 10$ ms and peak
luminosity is $\sim 10^{53}$ erg s${}^{-1}$.
This strong peak is illustrated as follows.
As a supernova shock moves outward, it dissociates nuclei into free
nucleons, which triggers the deleptonization process $e^-p\to\nu_en$;
these $\nu_e$ build up a sea because they are trapped and advected with
matter.
When the shock crosses the $\nu_e$ neutrinosphere, within which the
created $\nu_e$ are trapped, they are abruptly emitted.
For the other flavors $\bar\nu_e$ and $\nu_x$, there is no such sudden
burst; both luminosities glow rather gradually and they are similar to
each other.

The other characteristic that provides information on the flavor
conversion mechanism as well as neutrino decay is the hierarchy of the
average energy $\langle E_{\nu_e}\rangle < \langle E_{\bar\nu_e}\rangle
< \langle E_{\nu_x}\rangle$ as clearly seen from
Fig. \ref{fig:spectrum_org}; neutrino flavor conversion and decay also
change the spectral shape.
This energy hierarchy is explained as follows.
Since $\nu_x$ interacts with matter only through neutral-current
interactions in supernovae, they are weakly coupled with matter compared
to $\nu_e$ and $\bar\nu_e$.
Thus the neutrinosphere of $\nu_x$ is located deeper in the core than
that of $\nu_e$ and $\bar\nu_e$, which leads to higher temperatures for
$\nu_x$.
The difference between $\nu_e$ and $\bar\nu_e$ comes from the fact that
the core is neutron rich and the $\nu_e$ couples with matter more
strongly, through the $\nu_en\to e^-p$ reaction.

Although we only use one specific model \citep{Thompson03}, the
signature of the neutronization burst appears in all reliable numerical
simulations.
It is quite natural that the height and width of such a peak are
dependent on the supernova parameters as well as numerical approaches.
However, we are not interested in such a slight difference but focus
only on the rough signature---i.e., the {\it appearance or absence} of a
neutronization peak in the $\bar\nu_e$ channel.
We cannot discuss in any details without a concrete and reliable
original model, but even such rough information, if ever detected, would
bring a very fruitful and novel perspective to particle physics.
As for the neutrino spectrum, there is a general tendency of the
hierarchy of the average energies---i.e., $\langle E_{\nu_e}\rangle <
\langle E_{\bar\nu_e}\rangle < \langle E_{\nu_x}\rangle$ as already
noted and this tendency seems to be common to almost all numerical
calculations.
However, the spectral shape and especially the average energy ratio
between $\bar\nu_e$ and $\nu_x$ are still matter of controversy, and we
cannot conclude that the spectral information at the detectors would be
very useful at present (see, for a comparison among several
calculations, Ref. \citep{Keil03}).
Contrary to the luminosity curve, the neutrino spectrum does not
indicate such a prominent signature as a neutronization burst, and
therefore, the obtained spectrum would be of secondary importance for
the purpose of this study.
In the near future, however, the situation may become significantly
better, especially if some process relevant for successful supernova
explosions were discovered in computers; in that case, the spectrum as
well as luminosity curve would be very useful to precisely obtain the
unknown property of neutrinos.

Before moving on to a discussion including neutrino decay, here we
shortly describe the flavor conversions inside the supernova envelope
{\it without} any other processes such as decay.
Neutrinos of different mass eigenstates are expected to be incoherent
with each other when they reach the detector, and then the number
intensity (i.e., number per area, time, energy, and solid angle) of
$\bar\nu_e$ can be simply represented by
\begin{eqnarray}
 I_{\bar\nu_e}(L)&=&|U_{e1}|^2I_{\bar\nu_1}(R_{\rm SN})
  +|U_{e2}|^2I_{\bar\nu_2}(R_{\rm SN})\nonumber\\
 &&{}+|U_{e3}|^2I_{\bar\nu_3}(R_{\rm SN})\nonumber\\
 &\simeq&\cos^2\theta_{12}I_{\bar\nu_1}(R_{\rm SN})
  +\sin^2\theta_{12}I_{\bar\nu_2}(R_{\rm SN}),
  \label{eq:neutrino intensity without decay}
\end{eqnarray}
where $L$ and $R_{\rm SN}$ represent the distance to and radius of the
supernova, respectively.
The second equality comes from the fact that the value of $\theta_{13}$
is constrained to be negligibly small from reactor neutrino experiments
\citep{Apollonio99} and $\theta_{23}\simeq\pi/4$; from the solar and
reactor neutrino observation the obtained value for $\theta_{12}$ is
$\cos^2\theta_{12}\simeq 0.7$ (LMA solution)
\citep{Fukuda02a,Eguchi03a}.
The intensity at the supernova surface $I_{\nu}(R_{\rm SN})$ reflects
the flavor conversions during propagation inside the supernova envelope.
Flavor conversions during neutrino propagation have been extensively
studied by many researchers (see, e.g., Ref. \citep{Dighe00}), but we
briefly summarize them here.
In the case of a normal mass hierarchy ($m_1\ll m_3$), on which we focus
in this study, the intensity of each mass eigenstate at the surface is
fairly well known to be
\begin{eqnarray}
 I_{\bar\nu_1}(R_{\rm SN})&=&I_{\bar\nu_e}(0),\nonumber\\
 I_{\bar\nu_2}(R_{\rm SN})&=&I_{\bar\nu_3}(R_{\rm SN})
  =I_{\nu_x}(0),
  \label{eq:Intensity for antineutrinos: pure flavor conversion}
\end{eqnarray}
where $I_{\nu}(0)$ represents the neutrino intensity of each flavor
eigenstate at the neutrinosphere, for which we use the results of
numerical simulation shown in Figs. \ref{fig:luminosity_curve_org}
and \ref{fig:spectrum_org}.
By a combination of Eqs. (\ref{eq:neutrino intensity without decay}) and
(\ref{eq:Intensity for antineutrinos: pure flavor conversion}), we can
see that owing to flavor conversion inside the supernova, about 30\% of
the detected $\bar\nu_e$ would originate from $\nu_x$ at production;
this would harden the obtained spectrum at the detectors.
On the other hand, for the neutrino sector, the final expression for the
intensity of $\nu_e$ is given by the same expression as
Eq. (\ref{eq:neutrino intensity without decay}) but with replacing
$\bar\nu$ by $\nu$.
Flavor conversions inside the supernova envelope are, this time, a
little bit complicated; the unknown parameter $\theta_{13}$ strongly
affects the results.
Instead of $\theta_{13}$, we rather use so-called ``flip probability''
at the higher resonance point $P_H$ \citep{Dighe00}, which equals 0 (1)
when $\sin^22\theta_{13}\agt 10^{-3}$ ($\sin^22\theta_{13}\alt
10^{-5}$).
The expressions corresponding to Eq. (\ref{eq:Intensity for
antineutrinos: pure flavor conversion}) for the neutrino sector are then
given by
\begin{eqnarray}
 I_{\nu_1}(R_{\rm SN})&=&I_{\nu_x}(0),\nonumber\\
 I_{\nu_2}(R_{\rm SN})&=&P_HI_{\nu_e}(0)+(1-P_H)I_{\nu_x}(0),\nonumber\\
 I_{\nu_3}(R_{\rm SN})&=&(1-P_H)I_{\nu_e}(0)+P_HI_{\nu_x}(0).
  \label{eq:Intensity for neutrinos: pure flavor conversion}
\end{eqnarray}
These expressions are necessary for estimating the neutrino flux in the
case of possible decay, since $\nu_i\to\bar\nu_j$ might occur.

\section{Formulation \label{sec:Formulation}}

In this section, we derive new formulation for the detected $\bar\nu_e$
flux, which includes both flavor conversion and decay during
propagation.
The $\bar\nu_e$ intensity at the detectors is represented by
\begin{eqnarray}
 I_{\bar\nu_e}(L,E)&=&|U_{e1}|^2I_{\bar\nu_1}(L,E)
  +|U_{e2}|^2I_{\bar\nu_2}(L,E)\nonumber\\
 &&{}+|U_{e3}|^2I_{\bar\nu_3}(L,E),
  \label{eq:neutrino intensity with decay}
\end{eqnarray}
which is similar to Eq. (\ref{eq:neutrino intensity without decay}), but
the intensity of the specific mass eigenstate is no longer conserved
during its propagation owing to decay, $I_{\bar\nu_i}(L)\neq
I_{\bar\nu_i}(R_{\rm SN})$.
Here and from this point on, we explicitly show the neutrino energy
$E$.
The effect of neutrino decay on the intensity of each mass eigenstate
$\bar\nu_i$ is included by adding the appearance and disappearance terms
to the transfer equation---i.e.,
\begin{eqnarray}
 \frac{dI_{\bar\nu_i}}{dr}&=&
  -\sum_{j<i}{\mit\Gamma}_{ji}(E)I_{\bar\nu_i}(r,E)\nonumber\\
 &&{}+\sum_{j>i}\int_E^\infty
  dE^\prime\left[\psi_c(E^\prime,E){\mit\Gamma}_{\bar\nu_j\to\bar\nu_i}
	    (E^\prime)I_{\bar\nu_j}(r,E^\prime)\right.\nonumber\\
 &&{}\left.+\psi_f(E^\prime,E){\mit\Gamma}_{\nu_j\to\bar\nu_i}(E^\prime)
	  I_{\nu_j}(r,E^\prime)\right],
 \label{eq:neutrino propagation in vacuum}
\end{eqnarray}
where we define ${\mit\Gamma}_{ji}={\mit\Gamma}_{\bar\nu_i\to\nu_j}
+{\mit\Gamma}_{\bar\nu_i\to\bar\nu_j}$, etc.
A similar formulation holds for the neutrino sector, although we do not
show it explicitly.
The first and second sums of Eq. (\ref{eq:neutrino propagation in
vacuum}) reflect the disappearance and appearance of $\bar\nu_i$,
respectively.
Fortunately, this set of formulations can be analytically integrated
from $R_{\rm SN}$ to $L$.
In the case of three-flavor context and normal mass hierarchy, the
solution to Eq. (\ref{eq:neutrino propagation in vacuum}) is given by
\begin{widetext}
\begin{eqnarray}
 I_{\bar\nu_1}(L,E)&=&
  I_{\bar\nu_1}(R_{\rm SN},E)+
  \int_E^\infty dE^\prime
  \left[\frac{1-e^{-({\mit\Gamma}_{13}(E^\prime)
   +{\mit\Gamma}_{23}(E^\prime))L}}
  {{\mit\Gamma}_{13}(E^\prime)+{\mit\Gamma}_{23}(E^\prime)}
 J_{3\to 1}(E^\prime,E)
 +\frac{1-e^{-{\mit\Gamma}_{12}(E^\prime)L}}{{\mit\Gamma}
  _{12}(E^\prime)}J_{2\to 1}(E^\prime,E) \right.\nonumber\\
  &&{} \left.+\int_{E^\prime}^\infty dE^{\prime\prime}
  \left(\frac{1-e^{-{\mit\Gamma}_{12}(E^\prime)L}}
   {{\mit\Gamma}_{12}(E^\prime)}
   -\frac{1-e^{-({\mit\Gamma}_{13}(E^{\prime\prime})+
   {\mit\Gamma}_{23}(E^{\prime\prime}))L}}
   {{\mit\Gamma}_{13}(E^{\prime\prime})+{\mit\Gamma}_{23}(E^{\prime\prime})}
 \right)J_{3\to 2\to 1}(E^{\prime\prime},E^\prime,E)\right],
 \label{eq:Intensity for nu1bar}\\
 I_{\bar\nu_2}(L,E)&=&e^{-{\mit\Gamma}_{12}(E)L}I_{\bar\nu_2}(R_{\rm SN},E)
  +\int_E^\infty dE^\prime
  \frac{e^{-{\mit\Gamma}_{12}(E)L}
  -e^{-({\mit\Gamma}_{13}(E^\prime)+{\mit\Gamma}_{23}(E^\prime))L}}
  {{\mit\Gamma}_{13}(E^\prime)+{\mit\Gamma}_{23}(E^\prime)
  -{\mit\Gamma}_{12}(E)}J_{3\to 2}(E^\prime,E),
 \label{eq:Intensity for nu2bar}\\
 I_{\bar\nu_3}(L,E)&=&e^{-({\mit\Gamma}_{13}(E)+{\mit\Gamma}_{23}(E))L}
  I_{\bar\nu_3}(R_{\rm SN},E),
  \label{eq:Intensity for nu3bar}
\end{eqnarray}
where
\begin{eqnarray}
 J_{3\to 2\to 1}(E^{\prime\prime},E^\prime,E)
  &=&\frac{1}{{\mit\Gamma}_{13}(E^{\prime\prime})
  +{\mit\Gamma}_{23}(E^{\prime\prime})-{\mit\Gamma}_{12}(E^\prime)}
  \nonumber\\
 &&{}\times\left[\psi_c(E^\prime,E){\mit\Gamma}
	    _{\bar\nu_2\to\bar\nu_1}(E^\prime)
   J_{3\to 2}(E^{\prime\prime},E^\prime)
   +\psi_f(E^\prime,E){\mit\Gamma}_{\nu_2\to\bar\nu_1}(E^\prime)
   \tilde{J}_{3\to 2}(E^{\prime\prime},E^\prime)\right],
 \label{eq:I 321}\\
 J_{i\to j}(E^\prime,E)&=&
 \psi_c(E^\prime,E){\mit\Gamma}_{\bar\nu_i\to\bar\nu_j}(E^\prime)
 I_{\bar\nu_i}(R_{\rm SN},E^\prime)
 +\psi_f(E^\prime,E){\mit\Gamma}_{\nu_i\to\bar\nu_j}(E^\prime)
 I_{\nu_i}(R_{\rm SN},E^\prime),
 \label{eq:I ij}\\
 \tilde{J}_{i\to j}(E^\prime,E)&=&
 \psi_c(E^\prime,E){\mit\Gamma}_{\nu_i\to\nu_j}(E^\prime)
 I_{\nu_i}(R_{\rm SN},E^\prime)
 +\psi_f(E^\prime,E){\mit\Gamma}_{\bar\nu_i\to\nu_j}(E^\prime)
 I_{\bar\nu_i}(R_{\rm SN},E^\prime).
 \label{eq:tilde I ij}
\end{eqnarray}
\end{widetext}
In these expressions, we have used the assumption that
${\mit\Gamma}_{ij} R_{\rm SN}\ll 1$; i.e., neutrinos never decay during
their propagation inside the supernova envelope.
With this assumption, the intensity at the stellar surface
$I_{\nu}(R_{\rm SN},E)$ is, also in this case, represented by
Eqs. (\ref{eq:Intensity for antineutrinos: pure flavor conversion}) and
(\ref{eq:Intensity for neutrinos: pure flavor conversion}).
Thus we obtain the intensity of $\bar\nu_e$ at the detector using
Eqs. (\ref{eq:neutrino intensity with decay}), (\ref{eq:Intensity for
nu1bar})--(\ref{eq:tilde I ij}), (\ref{eq:Intensity for antineutrinos:
pure flavor conversion}), and (\ref{eq:Intensity for neutrinos: pure
flavor conversion}).

Although we have given a quite general expression, we are rather
interested in the more specific case, in which the neutronization peak
appears in the $\bar\nu_e$ channel; this is realized when the neutrino
masses are strongly hierarchical as already discussed in
Secs.~\ref{sec:Models of decaying neutrinos} and \ref{sec:Original
supernova neutrino signal}.
In this case, we obtain ${\mit\Gamma}_{\bar\nu_i\to\nu_j}={\mit\Gamma}
_{\bar\nu_i\to\bar\nu_j}={\mit\Gamma}_{ji}/2$, etc., and the energy
distribution factions are given by Eq. (\ref{eq:energy distribution of
daughter neutrinos}).
From this point on, we use $\tau_{12}/m$, $\tau_{13}/m$, and
$\tau_{23}/m$ as free parameters, which are related to
${\mit\Gamma}_{ij}$ by ${\mit\Gamma}_{ij}(E)=m/(\tau_{ij}E)$.

\section{Results \label{sec:Results}}

\begin{figure}[htbp]
\begin{center}
\includegraphics[width=8cm]{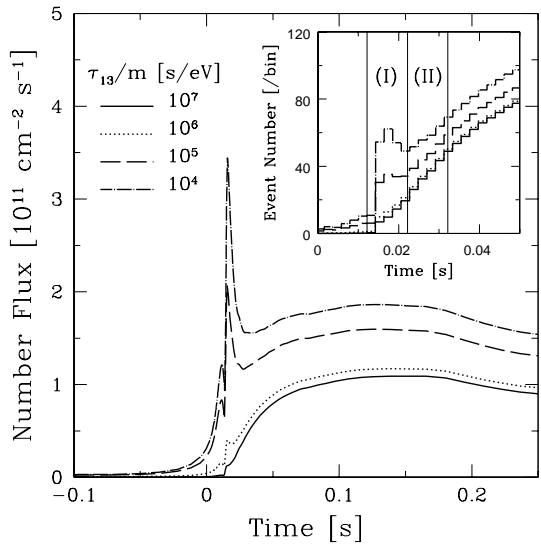}
\caption{Energy-integrated flux at the Earth in the case of adiabatic
 conversion $(P_H=0)$, for various values of $\tau_{13}/m$. The other
 parameters ($\tau_{12}/m$ and $\tau_{23}/m$) are set to infinity. The
 expected number of events at the detector of 640 kton is shown in the
 inset. \label{fig:luminosity_curve_AD}}
\end{center}
\end{figure}

\begin{figure}[htbp]
\begin{center}
\includegraphics[width=8cm]{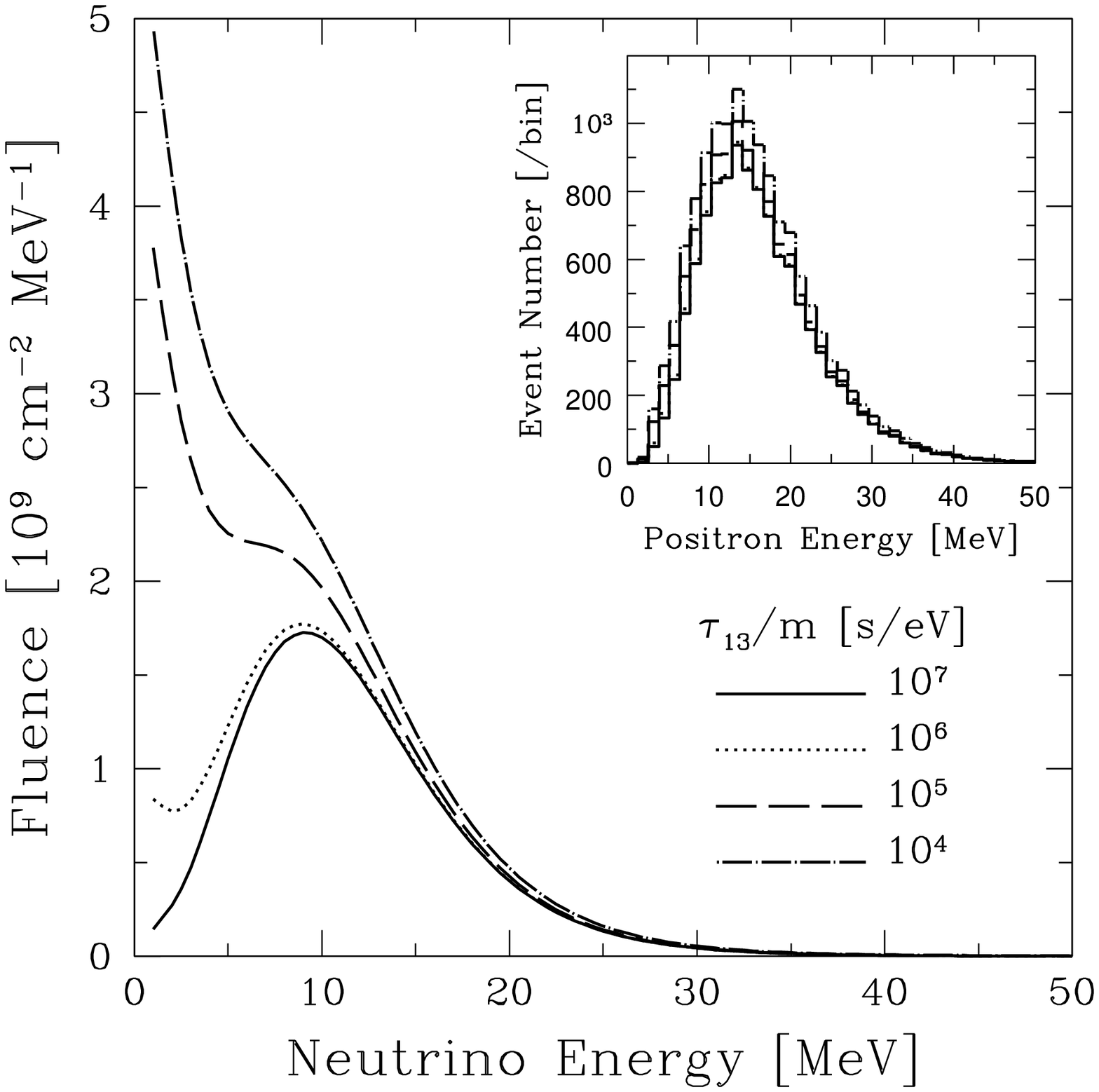}
\caption{Fluence (time-integrated flux) at the Earth in the case of
 adiabatic conversion $(P_H=0)$, for various values of
 $\tau_{13}/m$. The other parameters ($\tau_{12}/m$ and $\tau_{23}/m$)
 are set to infinity. The expected number of events at the detector of
 640 kton is shown in the inset.\label{fig:spectrum_AD}}
\end{center}
\end{figure}

\begin{figure}[htbp]
\begin{center}
\includegraphics[width=8cm]{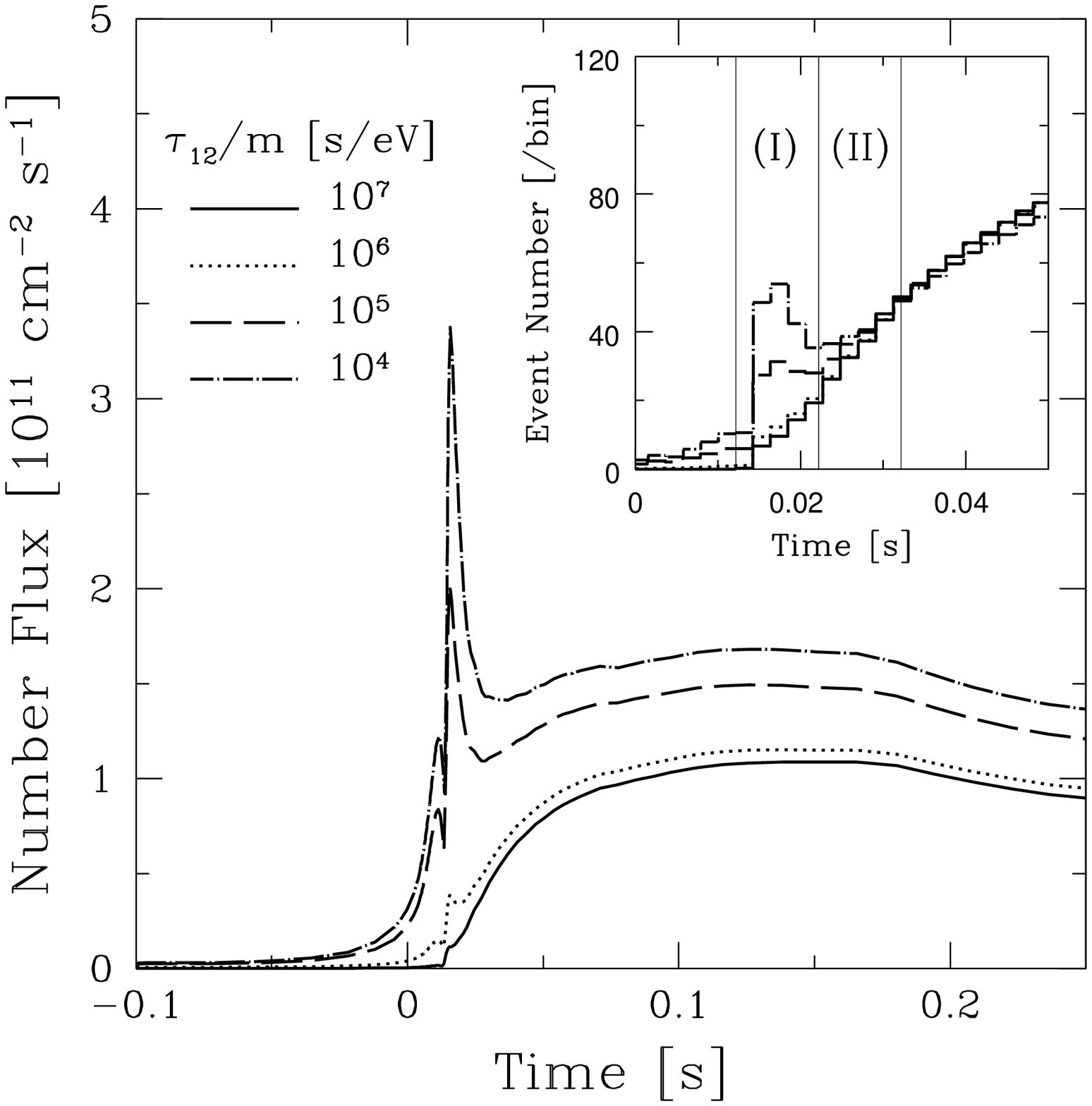}
\caption{The same as Fig. \ref{fig:luminosity_curve_AD} but for the
 case of nonadiabatic conversion ($P_H=1$), for various values of
 $\tau_{12}/m$ with fixed $\tau_{13}/m$ and
 $\tau_{23}/m$ to infinity. \label{fig:luminosity_curve_NAD}}
\end{center}
\end{figure}

\begin{figure}[htbp]
\begin{center}
\includegraphics[width=8cm]{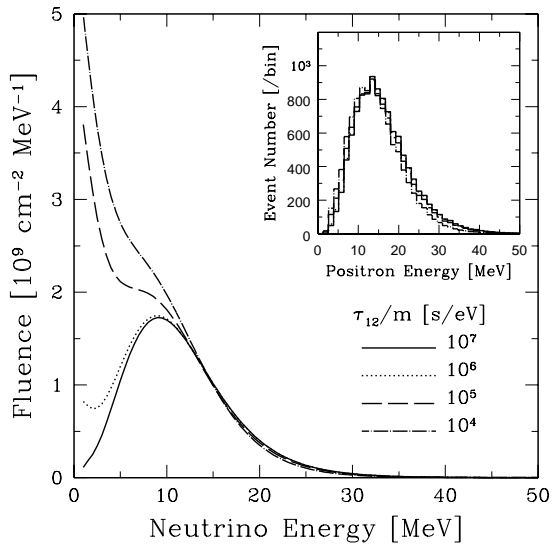}
\caption{The same as Fig. \ref{fig:spectrum_AD} but for the case of
 nonadiabatic conversion ($P_H=1$), for various values of $\tau_{12}/m$
 with fixed $\tau_{13}/m$ and $\tau_{23}/m$ to
 infinity. \label{fig:spectrum_NAD}}
\end{center}
\end{figure}

The obtained number flux and fluence (time-integrated flux) of
$\bar\nu_e$, using Eqs. (\ref{eq:neutrino intensity with decay}),
(\ref{eq:Intensity for nu1bar})--(\ref{eq:tilde I ij}),
(\ref{eq:Intensity for antineutrinos: pure flavor conversion}), and
(\ref{eq:Intensity for neutrinos: pure flavor conversion}) are shown in
Figs. \ref{fig:luminosity_curve_AD}--\ref{fig:spectrum_NAD}.
In the inset of each figure, we show the expected number of detection
at the water \v Cerenkov detectors with a fiducial volume of 640 kton
[20 times larger than that of the Super-Kamiokande (SK) detector], by
assuming that the supernova occurred at 10 kpc.
The values labeled in the vertical axis could be rescaled by using a
factor of $(10 ~{\rm kpc}/D)^2 (V_{\rm fid}/640 ~{\rm kton})$ in other
cases, where $D$ represents the distance and $V_{\rm fid}$ the fiducial
volume.
The cross section for the dominant catching process $\bar\nu_ep\to e^+n$
is fairly well understood and we used that given in
Ref. \citep{Vogel99}.
In addition we have used the trigger threshold expected to be installed
to SK-III (after full repair); with this threshold the
electrons and positrons of the energy of 3 MeV can be detected at 100\%
efficiency.
We neglect all other processes such as electron scattering because of
their subdominance.

Figures \ref{fig:luminosity_curve_AD} and \ref{fig:spectrum_AD} indicate
the energy- and time-integrated flux, respectively, in the case of the
adiabatic conversion, i.e., $P_H=0$.
The fluxes are evaluated for various values of $\tau_{13}/m$ with
$\tau_{12}/m$ and $\tau_{23}/m$ fixed to infinity.
The shape of the luminosity curve is found to strongly depend on the
lifetime of the $\nu_3(\bar\nu_3)\to\bar\nu_1$ mode.
This is because in the case of adiabatic conversion, the original
$\nu_e$ becomes $\nu_3$ at the stellar surface [Eq. (\ref{eq:Intensity
for neutrinos: pure flavor conversion})] and they decay into
$\bar\nu_1$, which dominantly couple to $\bar\nu_e$.
Thus, the peak of the neutronization burst clearly appears at the
detectors for $\tau_{13}/m < 10^{5}$ s/eV.
As for the spectrum, the energies of daughter neutrinos are
significantly degraded as shown in Fig. \ref{fig:spectrum_AD} and give a
very characteristic signature.
However, since the cross section is roughly proportional to $E^2$ and
highly insensitive to low-energy neutrinos, the expected event
number is almost the same at such a low-energy region as shown in the
inset of Fig. \ref{fig:spectrum_AD}.
This degradation of the neutrino energy due to its decay also causes the
actually detected neutronization peak to be less prominent compared to
that seen in the flux.

The case of nonadiabatic conversion ($P_H=1$) is shown correspondingly
in Figs. \ref{fig:luminosity_curve_NAD} and \ref{fig:spectrum_NAD}.
This time, the relevant parameter is changed to $\tau_{12}/m$ because
$\nu_e$ created by the neutronization is converted into $\nu_2$ at the
stellar surface.
The characteristics appearing in both the luminosity curve and spectrum
are similar to those in the case of adiabatic conversion, but the total
event number is slightly smaller.
This difference comes from the fact that the detected $\bar\nu_e$ is
coupled to $\bar\nu_2$ by $\sim 30$\% [Eq. (\ref{eq:neutrino intensity
with decay})] and the $\bar\nu_2$ disappears owing to its decay.
On the other hand, the disappearance of $\bar\nu_3$ does not directly
affect the expected event number since they hardly couple to
$\bar\nu_e$.

\begin{figure}[htbp]
\begin{center}
\includegraphics[width=8cm]{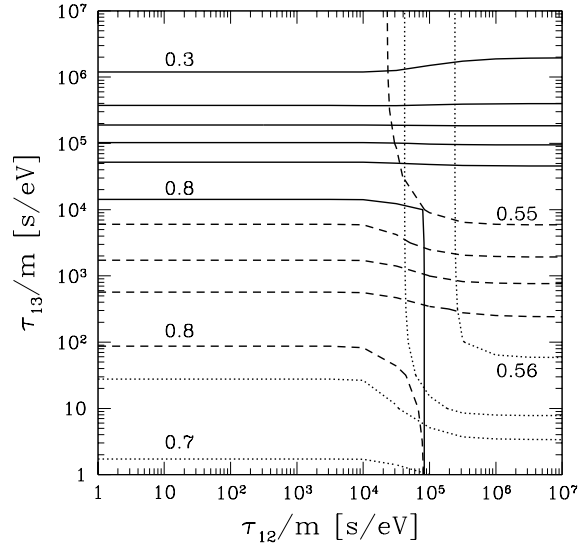}
\caption{Contour plot of $R_T$ against the ($\tau_{12}/m$, $\tau_{13}/m$)
 plane, in the case of the adiabatic conversion ($P_H=0$). The values of
 $\tau_{23}/m$ are taken to be $10^7$ (solid curves), $10^3$ (dashed
 curves), and $1$ (dotted curves) s/eV. Each curve of the same type is
 equally spaced by the value of $R_T$ with the indicated
 largest and smallest values. \label{fig:hardness_AD}}
\end{center}
\end{figure}

\begin{figure}[htbp]
\begin{center}
\includegraphics[width=8cm]{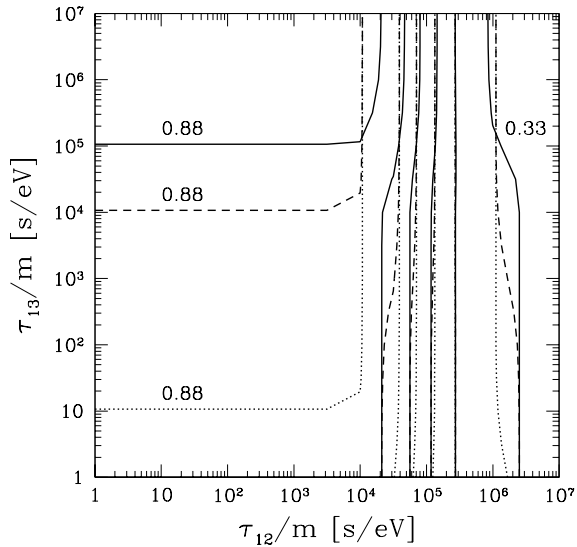}
\caption{The same as Fig. \ref{fig:hardness_AD} but for the case of the
 nonadiabatic conversion ($P_H=1$). \label{fig:hardness_NAD}}
\end{center}
\end{figure}

In order to discuss the parameter dependence of this mechanism, we
simply define the following quantity, which represents the peak
sharpness of the time profile.
Namely, it is defined as
\begin{equation}
 R_T=\frac{\mbox{event number in region (I)}}
  {\mbox{event number in region (II)}},
  \label{eq:R_T}
\end{equation}
where regions (I) and (II) are defined in the insets of
Figs. \ref{fig:luminosity_curve_AD} and \ref{fig:luminosity_curve_NAD}.
A larger value for $R_T$ means that the peak of the neutronization burst
is more prominent.
We plot the contour map of $R_T$ against the values of $\tau_{12}/m$ and
$\tau_{13}/m$ assuming several values for $\tau_{23}/m$; the result in
the case of adiabatic (nonadiabatic) conversion is shown in
Fig. \ref{fig:hardness_AD} (Fig. \ref{fig:hardness_NAD}).
In both figures, the solid, dashed, and dotted curves represent the case
that the values of $\tau_{23}/m$ are $10^{7}$ (essentially no decay due
to the $3\to 2$ mode), $10^3$, and $1$ s/eV, respectively.
Each curve of the same type is plotted at an equally spaced level,
while the largest and smallest values are indicated.
It can be clearly seen that in the case of $P_H=0$, the $R_T$ is
strongly dependent on the parameter $\tau_{13}/m$ but highly insensitive
to $\tau_{12}/m$.
This point also holds for the nonadiabatic case with the corresponding
replacement, $\tau_{13}\leftrightarrow\tau_{12}$.

\section{Discussion \label{sec:Discussion}}

In order to discuss how to discriminate one from the other decaying
models, in addition to $R_T$, we use another quantity that represents
the spectral hardness:
\begin{equation}
 R_{E}=\frac{\mbox{event number for $E_e>25$ MeV}}
  {\mbox{event number for $E_e<15$ MeV}}.
  \label{eq:R_E}
\end{equation}
The values for $R_E$ are obtained from the detected spectrum---i.e.,
from the insets of Figs. \ref{fig:spectrum_AD} and
\ref{fig:spectrum_NAD}.
As already discussed in Sec.~\ref{sec:Original supernova neutrino
signal}, because the average energy difference between each flavor
neutrino is still a matter of controversy, the obtained spectrum would
be also affected by such uncertainties.
We believe, however, that our treatment is quite reasonable, since using
the simple quantity $R_E$ would make the discussion rather insensitive
to such unsettled details.

\begin{figure}[htbp]
\begin{center}
\includegraphics[width=8cm]{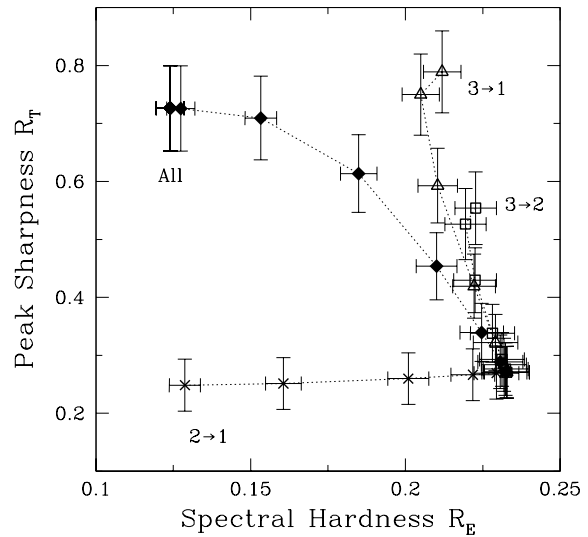}
\caption{Neutrino decay model plotted on the ($R_E,R_T$) plane in the
 case of adiabatic conversion. The error bars include only statistical
 errors, and are at the $1\sigma$ level, but their size should be
 accordingly rescaled as $(D/10~{\rm kpc})(V_{\rm fid}/640 ~{\rm kton})
 ^{-1/2}$. The labels represent the relevant mode (see text) and points
 of the same symbol indicate the model with a specific value of
 $\tau/m$, which is equally spaced from $10^4$ to $10^7$ s/eV,
 logarithmically. \label{fig:RRplot_AD}}
\end{center}
\end{figure}

Figure \ref{fig:RRplot_AD} shows the location of each model on the
($R_E,R_T$) plane for the adiabatic case.
We also show the $1\sigma$ statistical error bars of $R_E$ and $R_T$;
the size of these errors changes as $(D/10~{\rm kpc})(V_{\rm fid}/640
~{\rm kton})^{-1/2}$ in other cases.
Labels such as ``$2\to 1$'' represent the relevant decaying mode, while
the other modes are assumed to be stable; the label ``All'' represents
the case that $\tau_{12}/m = \tau_{13}/m = \tau_{23}/m$.
Points of the same symbol show how their location changes with lifetime;
each point of one symbol represents a model with a specific value of
lifetime-to-mass ratio $\tau/m$, which is equally spaced
(logarithmically) from $10^{4}$ to $10^{7}$ s/eV.
The dotted lines connect these points just to guide our eyes.
All the modes are degenerate when $\tau/m=10^{7}$ s/eV at $R_E=0.23$ and
$R_T=0.27$, which means that there occurs essentially no decay.
From this figure, models with extreme parameter values can be
distinguished from one another, if the currently planned megaton-class
detectors, such as Hyper-Kamiokande and UNO, detected the galactic
supernova neutrino burst.
On the other hand, in the case of currently working detectors such as
SK, the errors become very large by a factor of $\ge\sqrt{20}$, and
even using these very simple quantities $R_E$ and $R_T$, it would be
quite difficult to derive some information.

We here briefly illustrate the behavior of each model, shown in
Fig. \ref{fig:RRplot_AD}.
As already discussed in the previous section, the decaying mode from
$\nu_3(\bar\nu_3)$ to $\bar\nu_1$ makes the value of $R_T$ larger owing
to the appearance of a neutronization peak.
A similar explanation applies to the $3\to 2$ mode but its prominence is
reduced because the $\bar\nu_2$ state included in the $\bar\nu_e$ is
smaller compared to $\bar\nu_1$ state.
The $2\to 1$ mode does not change the peak sharpness $R_T$, since in the
case of adiabatic conversion, the $\nu_2 (\bar\nu_2)$ at the stellar
surface does not contain any component from the original $\nu_e$.
Instead of an almost constant $R_T$, the spectral hardness $R_E$
significantly changes with the value of $\tau_{12}/m$.
This is also easily illustrated as follows.
At the supernova surface, the $\nu_2$ and $\bar\nu_2$ both originate
from $\nu_x$, which shows the hardest spectral shape.
Without any decay, the obtained $\bar\nu_e$ signal then contains an
$\sim 30$\% amount of the original $\nu_x$.
On the contrary, if the decaying mode $2\to 1$ were relevant, the
original $\nu_x$ component, which otherwise should contribute to the
$\bar\nu_e$ spectrum, would disappear and instead it would appear as
$\bar\nu_1$ but with a significantly reduced energy; this makes the
spectral hardness $R_E$ considerably small.
The ``All'' model, in which we assumed $\tau_{12}/m = \tau_{13}/m =
\tau_{23}/m$, includes both effects given above; i.e., the peak
sharpness $R_T$ increases owing to the $3\to 1$ and $3\to 2$ modes,
while the spectral hardness $R_E$ decreases owing to the $2\to 1$ mode.

\begin{figure}[htbp]
\begin{center}
\includegraphics[width=8cm]{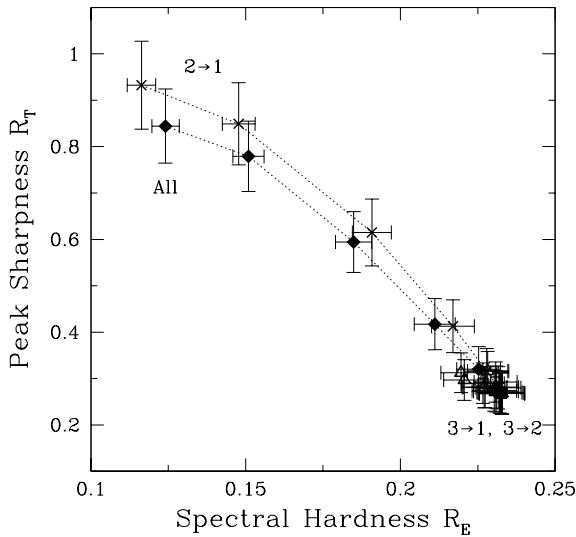}
\caption{The same as Fig. \ref{fig:RRplot_NAD}, but in the case of
 nonadiabatic conversion. \label{fig:RRplot_NAD}}
\end{center}
\end{figure}

Figure \ref{fig:RRplot_NAD} is the same as Fig. \ref{fig:RRplot_AD}, but
for the case of nonadiabatic conversion $P_H=1$.
In this case, the decaying mode from $\nu_3$ and $\bar\nu_3$ does not
essentially change the expected signal, because they are not coincident
with the original $\nu_e$ at the stellar surface as well as having
essentially no coupling to $\bar\nu_e$.
The original $\nu_e$, instead, in this case, appears as $\nu_2$; thus
the decaying mode $2\to 1$ considerably changes the detected signals.

\begin{figure}[htbp]
\begin{center}
\includegraphics[width=8cm]{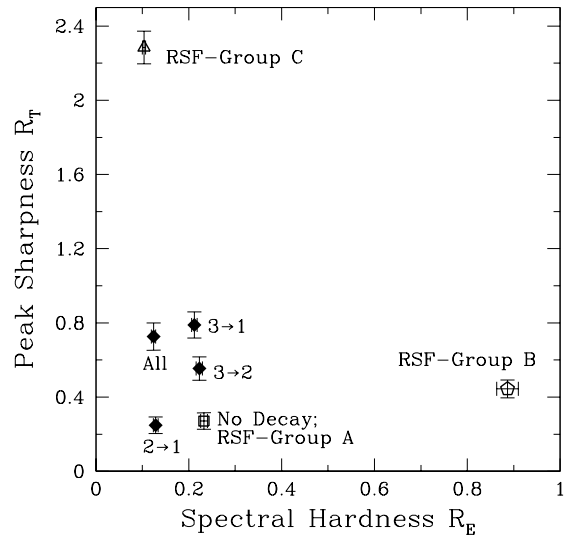}
\caption{Model groups A, B, and C due to the RSF mechanism given in
 Ref. \citep{Ando03g}, plotted on the ($R_{E},R_{T}$) plane. The
 decaying models for the adiabatic conversion with $\tau/m=10^4$ s/eV
 are also plotted for comparison. \label{fig:RRplot_RSF}}
\end{center}
\end{figure}

There also exists another mechanism that possibly changes the original
$\nu_e$ into a detected $\bar\nu_e$, resulting in the appearance of a
sharp peak due to a neutronization burst at the detectors---i.e.,
resonant spin-flavor (RSF) conversion \citep{Ando03g,Akhmedov03b}.
This mechanism is induced by the interaction between a supernova
magnetic field and the Majorana magnetic moment of neutrinos.
According to Ref. \citep{Ando03g}, the very sharp peak of a
neutronization burst could appear owing to the combination effect of the
RSF and ordinary matter-induced conversion, if the following conditions
are all satisfied: (i) the mass hierarchy is inverted, (ii) the value of
$\theta_{13}$ is sufficiently large, and (iii) the neutrino magnetic
moment as well as supernova magnetic field is large enough to induce the
adiabatic RSF conversion (but see also Ref. \citep{Akhmedov03b}).
In order to compare the RSF mechanism with the decaying models, we plot
the model groups given in Ref. \citep{Ando03g} in
Fig. \ref{fig:RRplot_RSF}.
For comparison we also plot the decaying models with $\tau/m=10^4$ s/eV
and $P_H=0$.
The models with $\tau/m=10^7$ s/eV (no-decay model) are degenerate with
group A due to the RSF conversion.
This figure clearly indicates that the RSF mechanism potentially gives a
far more characteristic $\bar\nu_e$ signal at the detectors; both the
peak sharpness of time profile $R_T$ (group C) and spectral hardness
$R_E$ (group B) are very prominent, and even the SK detector enables us
to discriminate these model groups.

\section{Conclusions \label{sec:Conclusions}}

Nonradiative decay of neutrinos is not constrained sufficiently; the
most stringent lower limit to the lifetime $\tau_{12}$ is obtained from
the solar neutrino observation but it is very weak \citep{Beacom02}.
Thus, neutrino decay possibly affects the neutrino signal from other
astrophysical objects such as supernovae.

Using newly derived formulation, which includes both flavor conversions
inside the supernova envelope and neutrino decays during propagation in
vacuum, we calculated the expected neutrino luminosity curve as well as
the spectrum at future large volume water \v Cerenkov detectors.
In these calculations, we particularly focused on the decaying model
such that the original $\nu_e$ appears in the $\bar\nu_e$ signal as
the result of flavor conversion and decay.
This is because such a situation can give the appearance of a sharp peak
in the time profile due to the neutronization burst, and it could be
easily recognized.
We discuss that this actually may be realized if the neutrino masses are
strongly hierarchical, and we have assumed it in actual calculations.
The lifetimes of three decaying modes, $\tau_{12}$, $\tau_{13}$, and
$\tau_{23}$, are taken to be free parameters, and the cases of adiabatic
and nonadiabatic conversions are treated independently.
The results of the calculations are shown in
Figs. \ref{fig:luminosity_curve_AD}--\ref{fig:spectrum_NAD} and the
neutronization peak can be significantly prominent in future
megaton-class water \v Cerenkov detectors.

In order to discuss the parameter dependence of the neutrino signal, we
introduce the rather simple quantities $R_T$ and $R_E$, which represent
the peak sharpness and spectral hardness, respectively.
As shown in Figs. \ref{fig:hardness_AD} and \ref{fig:hardness_NAD}, the
value of $R_T$ is strongly dependent on the value of $P_H$ as well as
the relevant lifetime.
From Figs. \ref{fig:RRplot_AD} and \ref{fig:RRplot_NAD}, we see that the
behaviors of each model on the ($R_E,R_T$) plane are considerably
different from one another.
But we also show that the location of these decay models on this plane
does not give prominent properties enough for us to distinguish using
current detectors such as SK, on the contrary to a very significant
dispersion due to the RSF conversion (Fig. \ref{fig:RRplot_RSF}).
Finally, we again stress that the appearance of the neutronization peak
clearly indicates nonstandard properties of neutrinos; neutrino decay
would then be one of the possible candidates.

\begin{acknowledgments}
This work was supported by a Grant-in-Aid from the JSPS.
\end{acknowledgments}


\bibliography{refs}

\end{document}